\documentclass
[superscriptaddress,secnumarabic,amssymb,amsmath,nobibnotes,aps,prd,showkeys,showpacs,nofootinbib,onecolumn,notitlepage]{revtex4}%
\usepackage{graphicx}
\usepackage{epsf}
\usepackage{bm}
\usepackage{amsmath}
\usepackage{amsfonts}
\usepackage{amssymb}
\usepackage{epstopdf}%
\providecommand{\U}[1]{\protect\rule{.1in}{.1in}}

\newcommand{\be}{\begin{equation}}
\newcommand{\ee}{\end{equation}}

\newcommand{\mincir}{\raise
-3.truept\hbox{\rlap{\hbox{$\sim$}}\raise4.truept\hbox{$<$}\ }}
\newcommand{\magcir}{\raise
-3.truept\hbox{\rlap{\hbox{$\sim$}}\raise4.truept\hbox{$>$}\ }}

\ifx\pdfoutput\relax\let\pdfoutput=\undefined\fi
\newcount\msipdfoutput
\ifx\pdfoutput\undefined\else
\ifcase\pdfoutput\else
\ifx\paperwidth\undefined\else
\ifdim\paperheight=0pt\relax\else\pdfpageheight\paperheight\fi
\ifdim\paperwidth=0pt\relax\else\pdfpagewidth\paperwidth\fi
\fi\fi\fi
\begin{document}
\title{Black branes in four-dimensional conformal equivalent theories}
\author{N. Dimakis}
\email{nsdimakis@gmail.com}
\affiliation{Center for Theoretical Physics, College of Physical Science and Technology Sichuan University, Chengdu 6100064, China}
\author{Alex Giacomini}
\email{alexgiacomini@uach.cl}
\affiliation{Instituto de Ciencias F\'{\i}sicas y Matem\'{a}ticas, Universidad Austral de
Chile, Valdivia, Chile}
\author{Andronikos Paliathanasis}
\email{anpaliat@phys.uoa.gr}
\affiliation{Instituto de Ciencias F\'{\i}sicas y Matem\'{a}ticas, Universidad Austral de
Chile, Valdivia, Chile}
\affiliation{Department of Mathematics and Natural Sciences, Core Curriculum Program,
Prince Mohammad Bin Fahd University, Al Khobar 31952, Kingdom of Saudi Arabia}
\affiliation{Institute of Systems Science, Durban University of Technology, PO Box 1334,
Durban 4000, Republic of South Africa}

\begin{abstract}
The physical properties of static analytic solutions which describe black brane
geometries are discussed. In particular we study the similarities and differences of analytic black brane/string solutions in the Einstein and Jordan frames. The comparison is made between vacuum power law $f(R)$ gravity solutions and their conformal equivalents in the Einstein frame. In our analysis we examine how the geometrical and physical
properties of these analytic axisymmetric solutions - such as
singularities, the temperature and the entropy - are affected as we pass from one frame to the other.

\end{abstract}
\date{\today}
\maketitle

\section{Introduction}

The need to determine a mechanism which provides a theoretical prediction on
the recent observable phenomena in cosmology, such as the late-time
acceleration phase of the universe \cite{dataacc1,dataacc2,data1,data2,data3},
has led cosmologists to introduce (alternative/modified) gravitational
theories which are defined in the Jordan frame \cite{Jord} instead of the
Einstein frame of General Relativity \cite{clifton1,mod0,mod1,mod2}.

One of the first proposed theories in the literature, which is formulated in
the Jordan frame, is the Brand-Dicke gravitational theory \cite{Brans}. More
specifically, Brans-Dicke theory is based on Mach's principle \cite{oha};
the gravitational field is described by the metric tensor and a scalar field
nonminimally coupled to gravity. Moreover, the gravitational constant $G$
depends on the value of nonminimally scalar field, and on a constant
$\omega_{BD}$ which is called Brans-Dicke parameter. Hence, $G$ is actually an
effective function rather a constant, for a discussion see also
\cite{jdb11,varcon}. In what regards the limits of parameter $\omega_{BD}$,
while someone expects General Relativity to be recovered when $\omega_{BD}$
reaches infinity, it was actually demonstrated that this is not true. This implies
that the two theories are fundamentally different \cite{omegabd}. On the other
hand, in the limit where the Brans-Dicke parameter vanishes, i.e. $\omega
_{BD}=0$, the theory is equivalent with that of a massive Dilaton theory
\cite{fujji}, as it has been shown by O'Hanlon\footnote{The Brans-Dicke action
with $\omega_{BD}=0$ is also known as O'Hanlon gravity.} \cite{oha2}.
 Moreover, Brans-Dicke theory provides a radiation epoch in the evolution of
the universe as also a de Sitter phase \cite{dyn1,dyn2}, which can be related
with the late-time acceleration phase of the universe \cite{ssen1}. It has
also been proposed that the Brans-Dicke parameter should not be a constant but
rather depend on the nonminimally coupled field. This led to a zoology of
theories known as Scalar-tensor theories \cite{far}. For other generalizations
of Brans-Dicke action see for instance
\cite{ref1,ref2,ref3,ref4,ref5,ref6,ref7,ref8,ref9,ref10}\ and references therein.

Scalar-tensor theories are conformally equivalent to General Relativity
\cite{far,con1,con2,Valf}. However the latter is just a mathematical equivalence
between the Jordan and the Einstein frame on the solution space of the field
equations. What is more, the latter is true only in the case where there are not any specific
conditions imposed from an extra matter source, that is except of the scalar
field; we refer the reader to a recent discussion in \cite{abd1}. In
particular, if there exists an additional matter source which is minimally
coupled to the scalar field in one of the frames, then under the conformal
transformation there will be an interaction between the matter source and the
scalar field, which means that the extra matter source will not be minimally
coupled to the latter. Moreover, while the energy conditions for the
scalar field are satisfied in the Einstein frame, they can be violated in the
Jordan frame \cite{mag1,mag2}. From cosmological studies, it has been found
that while a theory can provide an acceleration phase in one frame it may not
be able to do so in the other. From this it is clear the conformal equivalence
transformation between the two frames does not necessarily mean physical
equivalence; see also the discussions in
\cite{Valf,coneq2,coneq3,coneq4,coneq5,coneq6,coneq7,coneq8,Sanyal} and references
therein. The study of the physical properties between the Jordan and the
Einstein frame has been extended to astrophysical objects, such as black-holes
\cite{bhframe,bhframe2,bhframe3,bhframe4,bhframe5} or to even more exotic
configurations like the wormholes \cite{worm1,worm2,worm3} and the black
branes or strings \cite{bbcon1,bbcon2}.

There is a family of higher-order theories which can also be described in the form of second order
scalar-tensor theories \cite{fr1,fr0,fr01,fr2,fr3,fr4,fr5}. In these cases new degrees of freedom, which are introduced by the higher-order derivatives,
can be attributed to nonminimally scalar fields. The standard method to do
so is with the use of Lagrange multipliers \cite{lan1,lan2,lan3,lan4,lan5}.
Consequently, higher-order theories can provide the mechanics for the scalar
field description, an approach which has been applied to describe inflation \cite{star1,star2,star3,star4}.

One of the simplest higher-order theories which has drawn the attention of the
scientific society the last years is $f\left(  R\right)  $-gravity. More
specifically, it is a fourth-order theory in which the Einstein-Hilbert action
is modified so as to become a general function, $f$, of the Ricci scalar, $R$, of the
underlying geometry \cite{Buda}. Its equivalent description as a
scalar-tensor theory is in the form of Brans-Dicke theory with parameter
$\omega_{BD}~$zero. That is, $f\left(  R\right)  $-gravity is equivalent with
the so-called O'Hanlon theory \cite{fr1}. In the family of $f\left(  R\right)
$-gravity there belongs one of the most popular inflationary models, the
Starobinsky model for inflation \cite{star1}. There are other cosmological
tests that have shown that the theory can also describe the late-time
acceleration phase of the universe \cite{frlate0,frlate1,frlate2}. Some
astrophysical applications of $f\left(  R\right)  $-gravity can be found in
\cite{staa0,staa1,bh00,bh01,bh02,bh04,bh05,bh06,bh07,bh08} and a black hole topology theorem has been presented lately in \cite{Mishra}.

Recently, in the context of $f\left(  R\right)  $-gravity, a
new family of analytical solutions in a four-dimensional static spacetime was
found in \cite{nplb}. That new family of solutions can support black branes or strings in
vacuum when axisymmetry is introduced in the line element. These solutions are
not asymptotically maximally symmetric. The study of the thermodynamic
parameters for the black brane/string solution extracted in the context of the
power-law theory $f\left(  R\right)  =R^{k}$, has shown that the entropy
function has positive values only when $0<k<\frac{1}{2}$ or the power $k$ is a
negative, even integer.

The purpose of this work is to study the effects of the conformal
transformation between the Jordan and the Einstein frames for the black brane
solution determined in \cite{nplb}. Black branes/string solutions can be
easily constructed in Einstein's gravity and in some modified theories of
gravity by trivially embedding black hole solutions in higher dimensions
\cite{Kastor,Giribet,bs1,bs2,bs3,Chakra1,Chakra2}. However, the results of
\cite{nplb} are not based on this technique and the resulting spacetime is not constructed by the embedding of a lower dimensional solution in four dimensions. In this work we use the results obtained in \cite{nplb} so as to construct the analytical
solution of General Relativity with a minimally coupled scalar field for a
four-dimensional static axisymmetric spacetime. Furthermore, we shall study
the differences of the specific solutions; an important analysis in order the
investigate the physical (in-)equivalence of the two frames for static
axisymmetric solutions.

The plan of the paper is as follows: In Section \ref{sec2} we briefly discuss the conformal transformation which
connects the Jordan and the Einstein theory. In Section \ref{sec2a} we present the gravitational
field equations that correspond to the system under study. The analytical
solution which was determined in \cite{nplb} is discussed in Section
\ref{sec4}. The black brane solution in the Einstein frame is determined in Section \ref{sec5}. The
temperature and the entropy for the solution in the Einstein frame are
calculated in Section \ref{temp1}. Finally, in Section \ref{conc} we draw our conclusions.

\section{Conformal transformations}

\label{sec2}

By definition conformal transformations generalize the concept of isometries;
in particular, conformal transformations are angle-preserving transformations
compared to the isometries which preserve both the length and the angles between lines. Two metric tensors are conformally related when there exists a
scalar $\mathcal{N}\left(  x\right)  $ such that
\begin{equation}
\bar{g}_{\mu\nu}\left(  x\right)  =\mathit{\mathcal{N}}^{2}\left(  x\right)
g_{\mu\nu}\left(  x\right)  \label{con.01}%
\end{equation}
while the Ricci scalar $\bar{R}\left(  \bar{g}\right)  ~$of $\bar{g}_{\mu\nu}$
is given in terms of the Ricci scalar $R\left(  g\right)  $ for the metric
$g_{\mu\nu}$ and \textit{$\mathcal{N}$}$\left(  x\right)  $ as
\cite{hawellis}
\begin{equation}
\bar{R}\left(  \bar{g}\right)  =\mathit{\mathcal{N}}^{-2}R\left(  g\right)
-6\mathit{\mathcal{N}}^{-3}g_{\mu\nu}\mathit{\mathcal{N}}^{;\mu;\nu},
\label{con.02}%
\end{equation}
whenever $g_{\mu\nu}$~denotes a four-dimension space, that is, $\dim g_{\mu\nu}%
=4$. The covariant derivatives on the right hand side of \eqref{con.02} are defined with
respect to the metric $g_{\mu\nu}$. Two Action integrals, or more
specifically Lagrangians, are called conformally equivalent iff the solution
of the Euler-Lagrange equations survives under the action of the conformal
transformation \cite{abd1}.

Let us now consider the scalar-tensor theory in the four-dimensional space with
metric $\bar{g}_{\mu\nu}$, that is given by
\begin{equation}
S\left(  \varphi,\bar{g}\right)  =\int d^{4}x\sqrt{-\bar{g}}\left(  F\left(
\varphi\right)  \bar{R}\left(  \bar{g}\right)  -\frac{\alpha}{2}\bar{g}%
^{\mu\nu}\varphi_{;\mu}\varphi_{;\nu} - V\left(  \varphi\right)  \right)  ,
\label{lan.01}%
\end{equation}
where $\varphi$ denotes the nonminimally coupled scalar field and $V\left(
\varphi\right)  $ is the potential which drives the dynamics for this field. The parameter $\alpha$ is the analogue of the Brans-Dicke parameter,
$\alpha=\omega_{BD}$, when the function $F\left(  \varphi\right)  $ is
quadratic, that is\footnote{When $F\left(  \varphi\right)  \simeq\varphi^{2}$,
then under the reparametrizations $\varphi=\sqrt{\phi}$ for the scalar-field,
the Action integral (\ref{lan.01}) is written in the usual Brans-Dicke form \cite{Brans}.},
$F\left(  \phi\right)  \simeq\varphi^{2}$. On the other hand, when $\alpha=0$,
the Action integral (\ref{lan.01}) describes the O'Hanlon gravity which is
equivalent to the $f\left(  R\right)  $-gravity.

For the conformally related metric $g_{\mu\nu}$, the Action Integral
(\ref{lan.01}) with the use of \ expression (\ref{con.02}) becomes%
\begin{equation}
S\left(  \varphi,g\right)  =\int d^{4}x\sqrt{-g}\left(  F\left(
\varphi\right)  \mathit{\mathcal{N}}^{2}R\left(  g\right)  -\frac{\alpha}%
{2}\mathit{\mathcal{N}}^{2}g^{\mu\nu}\varphi_{;\mu}\varphi_{;\nu} -
\mathit{\mathcal{N}}^{4}V\left(  \varphi\right)  -6F\left(  \varphi\right)
\mathit{\mathcal{N}}g^{\mu\nu}\mathit{\mathcal{N}}_{;\mu\nu}\right)  .
\label{lan.03}%
\end{equation}
while integration by parts simplify the latter to \cite{kass}%
\begin{equation}
S\left(  \varphi,g\right)  =\int d^{4}x\sqrt{-g}\left(  F\left(
\varphi\right)  \mathit{\mathcal{N}}^{2}R\left(  g\right)  -\frac{\alpha}%
{2}\mathit{\mathcal{N}}^{2}g^{\mu\nu}\varphi_{;\mu}\varphi_{;\nu} -
\mathit{\mathcal{N}}^{4}V\left(  \varphi\right)  -6\mathit{\mathcal{N}%
}F_{,\varphi}g^{\mu\nu}\mathit{\mathcal{N}}_{;\mu}\phi_{;\nu}-6F\left(
\varphi\right)  g^{\mu\nu}\mathit{\mathcal{N}}_{;\mu}\mathit{\mathcal{N}%
}_{;\nu}\right)  . \label{lan.04}%
\end{equation}

It is straightforward to see that scalar \textit{$\mathcal{N}$}$\left(
x\right)  $ does not provide any new degrees of freedom and actually is a
degenerate function. Thus, without loss of generality, we can remove it from
the Action integral by choosing \textit{$\mathcal{N}$}$=$\textit{$\mathcal{N}%
$}$\left(  \phi\left(  x\right)  \right)  $. In such a case, the Action
Integral (\ref{lan.04}) is written as%
\begin{equation}
S\left(  \varphi,g\right)  =\int d^{4}x\sqrt{-g}\left(  \hat{F}\left(
\varphi\right)  R\left(  g\right)  -\frac{\alpha}{2}H\left(  \varphi\right)
g^{\mu\nu}\varphi_{;\mu}\varphi_{;\nu}-\hat{V}\left(  \varphi\right)  \right)
. \label{lan.05}%
\end{equation}
where
\begin{equation}
H\left(  \varphi\right)  =\left(  \mathit{\mathcal{N}}^{2}+\frac{12}{\alpha
}\left(  \mathit{\mathcal{N}}F_{,\varphi}\mathit{\mathcal{N}}_{,\varphi}+
F\left(  \varphi\right)  \left(  \mathit{\mathcal{N}}_{,\varphi}\right)  ^{2}
\right)  \right)  ,~~\hat{F}\left(  \varphi\right)  =F\left(  \varphi\right)
\mathit{\mathcal{N}}^{2},~\text{and}~\hat{V}\left(  \varphi\right)
=\mathit{\mathcal{N}}^{4}V\left(  \varphi\right)  ,
\end{equation}
or equivalently as follows
\begin{equation}
S\left(  \phi,R\right)  =\int d^{4}x\sqrt{-\gamma}\left(  \hat{F}\left(
\phi\right)  R\left(  g\right)  -\frac{\alpha}{2}g^{\mu\nu}\phi_{;\mu}%
\phi_{;\nu}- \hat{V}\left(  \phi\right)  \right)  \label{lan.007}%
\end{equation}
in which the scalar field $\phi\left(  x\right)  $ is related with
$\varphi\left(  x\right)  $ with the transformation \ $d\phi=\sqrt{H\left(
\phi\right)  }d\varphi$. Recall that the latter transformation is a coordinate
transformation and does not affect the dynamics of the system.

The Action Integral (\ref{lan.007}) still describes a scalar-tensor theory in
the Jordan frame. However, when $\hat{F}\left(  \phi\right)  =const.$, that is
when $F\left(  \varphi\right)  \mathcal{N}\left(  \varphi\right)  ^{2}%
=const.$, the Action Integral takes the form of the one of General Relativity
with a minimally coupled scalar field, i.e.
\begin{equation}
S\left(  \phi,R\right)  =\int d^{4}x\sqrt{-g}\left(  R\left(  g\right)
-\frac{1}{2}g^{\mu\nu}\phi_{;\mu}\phi_{;\nu} - \hat{V}\left(  \phi\right)
\right)  . \label{lan.008}%
\end{equation}
where without loss of generality we have chosen $\alpha=1$.

Indeed, the two Action integrals (\ref{lan.01}) and (\ref{lan.008}) are
conformally related but it is important to mention that $g$ and $\bar{g}$ are
in principle two different line elements. In the sense that there does not
necessarily exist a space-time diffeomorphism linking the two metrics. Hence,
while the above transformation can be used to map solutions between the two
frames, one should be very careful about the possible physical
(in-)equivalence of the two theories.

At this point, just for the convenience of the reader, we briefly want to
demonstrate the equivalence of $f\left(  R\right)  $-gravity with Brans-Dicke
theory with $\omega_{BD}=0$, that is the O'Hanlon theory. The Action Integral
in $f\left(  R\right)  $-gravity, for the line element $\bar{g}_{\mu\nu}$, is
given as follows
\begin{equation}
S_{f\left(  R\right)  }=\int d^{4}x\sqrt{-g}f\left(  \bar{R}\left(  \bar
{g}\right)  \right)  , \label{lan.09}%
\end{equation}
with the field equations resulting from the variation with respect to the
metric being
\begin{equation}
\label{fRfieldeq}f_{,R} R_{\mu\nu} - \frac{1}{2} f g_{\mu\nu}- \left(
\nabla_{\mu}\nabla_{\nu}- g_{\mu\nu} \nabla_{\sigma}\nabla^{\sigma}\right)
f_{R} = 0.
\end{equation}
Without loss of generality we introduce the Lagrange multiplier $\lambda$,
such that%
\begin{equation}
S_{f\left(  R\right)  }=\int d^{4}x\sqrt{-g}\left(  f\left(  \bar{R}\left(
\bar{g}\right)  \right)  +\lambda\left(  \bar{R}\left(  \bar{g}\right)
-\bar{R}_{ex}\right)  \right)  . \label{lan.10}%
\end{equation}
where $\bar{R}_{ex}$ denotes the expansion of the Ricci scalar in terms of the
functions of the metric $\bar{g}_{\mu\nu}$. Variation with respect to the
Ricci scalar $\bar{R}$ at the Action Integral (\ref{lan.10}) gives the Euler-Lagrange equation $\frac{\delta S_{f\left(  R\right)  }}{\delta\bar{R}}=0$, or
$\lambda=-f_{,\bar{R}}\left(  \bar{R}\right)  $, which defines the Lagrange
multiplier $\lambda$. By substituting the latter in (\ref{lan.10}) it follows
that \cite{fr5}%
\begin{equation}
S_{f\left(  R\right)  }=\int d^{4}x\sqrt{-g}\left(  f_{,\bar{R}}\left(
\bar{R}\right)  \bar{R}_{ex}+\left(  f\left(  \bar{R}\right)  -f_{,\bar{R}%
}\bar{R}\right)  \right)  \label{lan.11}%
\end{equation}
and by defining the new scalar $\varphi=f_{,\bar{R}}\left(  \bar{R}\right)
$ we obtain the scalar-tensor theory%
\begin{equation}
S_{f\left(  R\right)  ,\phi}=\int d^{4}x\sqrt{-g}\left(  \varphi\bar{R}%
_{ex}\left(  \bar{g}\right)  - V_{f}\left(  \varphi\right)  \right),
\label{lan.12}%
\end{equation}
in which $V_{f}\left(  \varphi\right)  =\left(  \varphi\bar{R}\left(
\varphi\right)  - f\left(  \bar{R}\left(  \varphi\right)  \right)  \right)  $.
This latter action\footnote{Recall that by definition $\bar{R}_{ex}\equiv
\bar{R}\left(  \bar{g}\right)  .$} describes the Brans-Dicke theory with
$\omega_{BD}=0$. The corresponding equations for the gravitation and scalar field are
\begin{subequations}
\label{OHfieldeq}%
\begin{align}
&  \varphi\left(  R_{\mu\nu}-\frac{1}{2}g_{\mu\nu} R \right)  + \left(
g_{\mu\nu}\nabla_{\sigma}\nabla^{\sigma}- \nabla_{\mu}\nabla_{\nu}\right)
\varphi+ \frac{1}{2} g_{\mu\nu} V(\varphi) = 0\\
&  R - V^{\prime}(\varphi) =0
\end{align}
respectively. It is also true that
$f\left(  R\right)  $-gravity defined in the Jordan frame is conformally
equivalent to General Relativity induced by a minimally coupled scalar field.
The conformal transformation which relates the Action Integrals (\ref{lan.008}%
), (\ref{lan.12}) is \cite{fr1}
\end{subequations}
\begin{equation}
g_{\mu\nu}=\varphi\bar{g}_{\mu\nu}~~,~\phi=\sqrt{3}\ln\varphi. \label{lan.14}%
\end{equation}

In the following section, we proceed with the derivation of the field
equations for the static axisymmetric spacetime in the Einstein and Jordan
frames, in particular that of $f\left(  R\right)  -$gravity. Moreover, we
assume that there is not any extra matter source in the gravitational system
except of the scalar field.

\section{Field equations in static axisymmetric spacetime}

\label{sec2a}

Consider the four-dimensional static axisymmetric spacetime with line element%
\begin{equation}
d\bar{s}^{2}=-a(r)^{2}dt^{2}+n(r)^{2}dr^{2}+b(r)^{2}\left(  dy^{2}%
+dz^{2}\right)  , \label{met.01}%
\end{equation}
which in general admits a four dimensional Killing algebra, consisted by the
vector fields $\left\{  \partial_{t},~\partial_{y},~\partial z,~z\partial
_{y}-y\partial_{z}\right\}  $.

In $f\left(  R\right)  $-gravity, the gravitational field equations
\eqref{fRfieldeq} for metric (\ref{met.01}) are equivalent to those produced
by the variation, with respect to the variables $\left\{  N\left(  r\right)
,a\left(  r\right)  ,b\left(  r\right)  ,R\left(  r\right)  \right\}  $, of
the Lagrangian function%
\begin{equation}
\mathcal{L}_{f\left(  R\right)  }=\frac{2}{n}\left[  f_{,R}\left(
ab^{\prime2}+ 2 ba^{\prime}b^{\prime}\right)  +f_{,RR}\left(  b^{2}a^{\prime
}R^{\prime}+2abb^{\prime}R^{\prime}\right)  \right]  +n a b^{2}\left(
f-Rf_{,R}\right)  , \label{met.02}%
\end{equation}
where the prime denotes a total
derivative with respect to the variable $r$, that is $x^{\prime}\left(
r\right)  =\frac{dx\left(  r\right)  }{dr}$.

On the other hand, in the scalar-tensor equivalent description, Lagrangian
(\ref{met.02}) takes the equivalent form%
\begin{equation}
\mathcal{L}_{\varphi}=\frac{2}{n}\left[  \varphi\left(  ab^{\prime
2}+2ba^{\prime}b^{\prime}\right)  +\left(  b^{2}a^{\prime}\varphi^{\prime
}+2abb^{\prime}\varphi^{\prime}\right)  \right]  - n a b^{2}V\left(
\varphi\right)  , \label{met.03}%
\end{equation}
while the corresponding Euler - Lagrange equations yield
\begin{equation}
\frac{2}{n^{2}}\left[  \varphi\left(  ab^{\prime2}+2ba^{\prime}b^{\prime
}\right)  +\left(  b^{2}a^{\prime}\varphi^{\prime}+2abb^{\prime}%
\varphi^{\prime}\right)  \right]  + ab^{2}V\left(  \varphi\right)  =0,
\label{met.04}%
\end{equation}%
\begin{equation}
6a^{\prime\prime}=\frac{2a}{b^{2}}b^{\prime2}+\frac{6}{n}a^{\prime}n^{\prime}
- an^{2}V_{,\varphi}-\frac{4}{\varphi}a^{\prime}\varphi^{\prime}%
+\frac{4b^{\prime}}{b\varphi}\left(  a\varphi^{\prime}-2\varphi a^{\prime
}\right)  , \label{met.05}%
\end{equation}%
\begin{equation}
6b^{\prime\prime}=-\frac{4}{b}b^{\prime2} - bn^{2}V_{,\phi}+2b^{\prime}\left(
3\frac{n^{\prime}}{n}-\frac{\varphi^{\prime}}{\varphi}\right)  +\frac
{2}{a\varphi}a^{\prime}\left(  b\phi^{\prime}-\phi b^{\prime}\right)  ,
\label{met.06}%
\end{equation}%
\begin{equation}
6\varphi^{\prime\prime}=n^{2}\left(  2\varphi V_{,\varphi} -3V\left(
\varphi\right)  \right)  +\frac{6}{n}n^{\prime}\varphi^{\prime}+\frac{2\varphi
b^{\prime}\left(  2ba^{\prime}+ab^{\prime}\right)  -4b\left(  ba^{\prime
}+2ab^{\prime}\right)  \phi^{\prime}}{ab^{2}}. \label{met.07}%
\end{equation}
It can be straightforwardly verified that the above system is also equivalent
to field equations \eqref{OHfieldeq} under the ansatz \eqref{met.01} for the
line element.

The conformal transformation (\ref{lan.14}) reads%
\begin{equation}
a\left(  r\right)  \rightarrow\varphi\left(  r\right)  ^{-\frac{1}{2}}A\left(
r\right)  ~,~b\left(  r\right)  \rightarrow\varphi\left(  r\right)
^{-\frac{1}{2}}A\left(  r\right)  ~,~n\left(  r\right)  \rightarrow
\varphi\left(  r\right)  ^{-\frac{1}{2}}N\left(  r\right)  ~,~\varphi
\rightarrow e^{\frac{\sqrt{3}}{3}\phi} \label{met.08}%
\end{equation}
and under its use the field equations (\ref{met.04})-(\ref{met.07}) become%
\begin{equation}
\frac{1}{N^{2}}\left(  2 AB^{\prime2}+ 4 BA^{\prime}B^{\prime}-\frac{1}%
{2}AB^{2}\phi^{\prime2}\right)  + AB^{2}N\hat{V}\left(  \phi\right)  =0,
\label{met.09}%
\end{equation}
\qquad\qquad%
\begin{equation}
8A^{\prime\prime}=8A^{\prime}\left(  \frac{N^{\prime}}{N}-\frac{B^{\prime}}%
{B}\right)  +4\frac{A}{B^{2}}B^{\prime2}- A\phi^{\prime2} -2 A N^{2}\hat
{V}\left(  \phi\right)  , \label{met.10}%
\end{equation}%
\begin{equation}
8B^{\prime\prime}=-\frac{4}{B}B^{\prime2}+\frac{8}{N}N^{\prime}B^{\prime}-
B\phi^{\prime2}- 2 B N^{2}\hat{V}\left(  \phi\right)  , \label{met.11}%
\end{equation}
and%
\begin{equation}
\phi^{\prime\prime}=-\left(  \frac{A^{\prime}}{A}+2\frac{B^{\prime}}%
{B}\right)  \phi^{\prime}+\frac{N^{\prime}}{N}\phi^{\prime} + N^{2}\hat
{V}_{,\phi}, \label{met.12}%
\end{equation}
in which $\hat{V}\left(  \phi\right)  =e^{-\frac{2\sqrt{3}}{3}\phi}V\left(
\phi\right)  $. Finally, under the transformation (\ref{met.08}) the line
element (\ref{met.01}) can be written as $d\bar{s}^{2}=\varphi^{-1} ds^{2}$,
where the new line element $ds^{2}$ is now
\begin{equation}
ds^{2}=-A(r)^{2}dt^{2}+n(r)^{2}dr^{2}+B(r)^{2}\left(  dy^{2}+dz^{2}\right)  .
\label{met.14}%
\end{equation}

The set of equations (\ref{met.09})-(\ref{met.12}) describe the Einstein field
equations with a minimally coupled scalar field, i.e. those generated from
Action Integral (\ref{lan.008}), but for the line element (\ref{met.14}),
which is different from that of (\ref{met.01}). Although the solutions of
the conformal equivalent theories commute thought the conformal
transformation (\ref{met.08}), the physical and geometrical properties of the solutions can
be extremely different due fact that a conformal transformation cannot in general correspond to a diffeomorphism linking the two line elements.

\section{Black branes in the Jordan frame}

\label{sec4}

In this section, we briefly discuss the analytical solution for the
gravitational field equations (\ref{met.04})-(\ref{met.07}) of $f\left(
R\right)  $-gravity which was derived recently in \cite{nplb}, and can describe
four-dimensional black branes (or strings depending on the topology on which
one endows the two dimensional surface). It was found in \cite{nplb} that the
line elements
\begin{equation}
d\bar{s}^{2}=-r^{2}\left(  \frac{r^{\frac{5-4k}{k-1}}}{m}-1\right)
dt^{2}+r^{\frac{-4k^{2}+4k+2}{(k-1)(2k-1)}}\left(  \frac{r^{\frac{5-4k}{k-1}}%
}{m}-1\right)  ^{-1}dr^{2}+r^{2}\left(  dy^{2}+dz^{2}\right)  ,~\text{for
}1<k<\frac{5}{4} \label{met.16}%
\end{equation}
and%
\begin{equation}
d\bar{s}^{2}=-r^{2}\left(  1-\frac{r^{\frac{5-4k}{k-1}}}{m}\right)
dt^{2}+r^{\frac{-4k^{2}+4k+2}{(k-1)(2k-1)}}\left(  1-\frac{r^{\frac{5-4k}%
{k-1}}}{m}\right)  ^{-1}dr^{2}+r^{2}\left(  dy^{2}+dz^{2}\right)  ~,~\text{for
}k<1~\text{or~}k>\frac{5}{4},~k\neq\frac{1}{2}. \label{met.17}%
\end{equation}
solve field equations \eqref{fRfieldeq} for a power-law $f\left(  R\right)
=R^{k}$ theory, with $k\neq0,1, \frac{1}{2},\frac{5}{4}$. Given the form of
the transformation that maps $f(R)$-gravity to the O'Hanlon theory we expect
that the corresponding potential $V_{f}\left(  \varphi\right)  $ also has a
power law form. Specifically, we can see that the aforementioned potential is
\begin{equation}
V_{f}\left(  \varphi\right)  = \pm\frac{6 (4 k-5) }{2 k-1} \varphi^{\frac
{k}{k-1}}. \label{met.15}%
\end{equation}
and the corresponding scalar field is given by
\begin{equation}
\label{OHf}\varphi(r) = r^{\frac{2 (k-2)}{2 k-1}} .
\end{equation}
Expressions \eqref{met.15} and \eqref{OHf} satisfy the field equations
\eqref{OHfieldeq} for line element \eqref{met.16} when we have the plus sign
in \eqref{met.15}. On the contrary, when we consider line element
\eqref{met.17}, we need to take into account the minus sign in potential \eqref{met.15}.

The gravitational properties of the solutions in $f(R)$-gravity and the
O'Hanlon theory are identical since they refer to the same metrics. For these
two line elements the Ricci scalar and the Kretschmann scalar are derived to
be
\begin{equation}
R=\pm\frac{6k(4k-5)r^{\frac{2(k-2)}{(k-1)(2k-1)}}}{(k-1)(2k-1)},
\label{met.18}%
\end{equation}
and\qquad%
\begin{align}
K  &  =\frac{4}{(1-2k)^{2}(k-1)^{2}m^{2}}\Big[\left(  56k^{4}-264k^{3}%
+468k^{2}-370k+111\right)  r^{-\frac{2\left(  8k^{2}-16k+9\right)
}{(2k-1)(k-1)}}+\nonumber\\
&  -2m(k-2)^{2}(2k-1)r^{\frac{-8k^{2}+18k-13}{(2k-1)(k-1)}}+3m^{2}\left(
8k^{4}-20k^{3}+13k^{2}-2k+2\right)  r^{\frac{4(k-2)}{(k-1)(2k-1)}}\Big].
\label{met.19a}%
\end{align}
From the latter expressions it is clear that the spacetimes admit curvature
singularity at $r=0$, when $k<1/2~$and $1<k\leq2.~$Moreover, \ it is
straightforward to observe that a coordinate singularity exists at
$r=r_{h}=m^{\frac{1-k}{4k-5}}$, when $m>0$ \cite{nplb}. It is for these values
of the parameters that we can consider that the space-time describes a black
brane or a black string (if in this latter case we consider $x$ and $y$ as
periodic variables). While the resulting spacetime is asymptotically Ricci
flat, we can see that the curvature tensor $R_{\kappa\lambda\mu\nu}$ does not
tend to zero as $r$ reaches infinity, which means that the spacetime cannot be
asymptotically deformed to a maximally symmetric space. For a discussion on
this kind of topological black holes (especially for a toroidal case
two-surface) we refer the reader to \cite{Vanzo}.

In this context we can also study the type of an effective fluid corresponds
two the line elements \eqref{met.16} and \eqref{met.17}. By considering the
anisotropic energy-momentum tensor $T^{(\varphi)}_{\mu\nu}= R_{\mu\nu}%
-\frac{1}{2}g_{\mu\nu} R$ generated by these geometries we may derive the
energy density to be
\begin{equation}
\rho_{\varphi}\left(  r\right)  = \pm\frac{r^{\frac{2 (k-2)}{(k-1) (2 k-1)}}%
}{m\left(  2k-1\right)  \left(  k-1\right)  }\left(  r^{\frac{5-4k}{k-1}%
}\left(  2k-3\right)  \left(  k-2\right)  +m\left(  6k^{2}-7k-1\right)
\right)  , \label{met.20}%
\end{equation}
where the plus sign corresponds to metric \eqref{met.16} and the minus to
\eqref{met.17} (together with the appropriate range of values for $k$). For
the case where $1<k<\frac{5}{4}$ we can see that $\rho_{\varphi}$ is positive
as long as $r> A_{k} r_{h}$, where
\begin{equation}
\label{seqA}A_{k} = \left(  \frac{2 k^{2}-7 k+6}{-6 k^{2}+7 k+1}\right)
^{\frac{k-1}{4 k-5}} .
\end{equation}
The multiplicative constant $A_{k}$ in front of the horizon radius $r_{h}$
assumes values in the segment $(1,e)$ as $k$ varies from $1$ to $\frac{5}{4}$.
In the other range for $k$ corresponding to solution \eqref{met.17} the
situation is more complicated. In order to have a positive energy density
(\ref{met.20}) one of the following conditions has to be met:

\begin{itemize}
\item $\frac{1}{2}<k<1$, with $r< A_{k} r_{h}$

\item $\frac{1}{12} \left(  7-\sqrt{73}\right)  <k<\frac{1}{2}$ or $\frac
{5}{4}<k<\frac{1}{12} \left(  \sqrt{73}+7\right)  $, with $r>A_{k} r_{h}$

\item $\frac{3}{2}<k<2$, with $r< A_{k} r_{h}$.
\end{itemize}

A positive energy density in a region that rests outside the horizon is only
possible in the second of the above cases.

The effective energy-momentum tensor is anisotropic. The radial pressure
$p^{\varphi}_{r}\left(  r\right)  $ in the $r$ direction is given by
\begin{equation}
p^{\varphi}_{r}\left(  r\right)  = \pm\frac{r^{\frac{-8 k^{2}+16 k-9}{(k-1) (2
k-1)}}}{(1-k) m} \left(  3 (k-1) m r^{\frac{4 k-5}{k-1}}+k-2\right)  ,
\label{met.21}%
\end{equation}
while the isotropic tangential pressure in the $x$ and $y$ directions is
\begin{equation}
p^{\varphi}_{T}=p^{\varphi}_{x}\left(  r\right)  =p^{\varphi}_{y}\left(
r\right)  = \pm\frac{r^{\frac{2 (k-2)}{(k-1) (2 k-1)}}}{(2 k-1)(k-1) m}
\left(  \left(  -6 k^{2}+7 k+1\right)  m+2 (k-2) (k-1) r^{\frac{5-4 k}{k-1}%
}\right)  .
\end{equation}
We can define the mean pressure $p_{\varphi} = \frac{1}{3}\left(  p^{\varphi
}_{r}+p^{\varphi}_{x}+p^{\varphi}_{y}\right)  $ and with respect to the latter
write a ``mean" equation of state parameter $w_{_{\varphi}}\left(  r\right)
=\frac{p_{_{\varphi}}\left(  r\right)  }{\rho_{_{\varphi}}\left(  r\right)  }$.

What we may observe is that near the curvature singularity $r\rightarrow0$ and
for the values of $k$ corresponding to solution \eqref{met.17} the equation of
state parameter has the limit $w_{_{\varphi}}\left(  r\rightarrow0\right)
=\frac{1}{3}$. That \textquotedblleft anisotropic\textquotedblright\ radiation
fluid is provided by the extra terms of $f\left(  R\right)  $-gravity
\cite{amendola}. We can now proceed with the study of the analytical solution
in the Einstein frame.

\section{Black branes in the Einstein frame}

\label{sec5}

At this point we can map the results expressed in the previous section to the
Einstein frame by performing transformation \eqref{lan.14}. After a
reparametrization with respect to the $r$ variable\footnote{The $r$ appearing
in this section is not the same as the one appearing in the previous.} the
resulting spacetime becomes
\begin{equation}
\label{einfr1}ds^{2} = - r^{2} \left(  \frac{r^{-\frac{(2 k-1) (4 k-5)}{3
(k-1)^{2}}}}{m}-1\right)  dt^{2} + \left(  \frac{1-2 k}{3 (k-1)}\right)  ^{2}
\frac{r^{\frac{-4 k^{2}+4 k+2}{3 (k-1)^{2}}}}{\frac{r^{-\frac{(2 k-1) (4
k-5)}{3 (k-1)^{2}}}}{m}-1} +r^{2}\left(  dy^{2}+dz^{2}\right)  ,
\end{equation}
for $1<k<\frac{5}{4} $ or
\begin{equation}
\label{einfr2}ds^{2} = - r^{2} \left(  1-\frac{r^{-\frac{(2 k-1) (4 k-5)}{3
(k-1)^{2}}}}{m}\right)  dt^{2} + \left(  \frac{1-2 k}{3 (k-1)}\right)  ^{2}
\frac{r^{\frac{-4 k^{2}+4 k+2}{3 (k-1)^{2}}}}{1-\frac{ r^{-\frac{(2 k-1) (4
k-5)}{3 (k-1)^{2}}}}{m}} +r^{2}\left(  dy^{2}+dz^{2}\right)
\end{equation}
for $k \in(-\infty, 1)\bigcup(\frac{5}{4},+\infty)- \{\frac{1}{2}\}$. The
corresponding minimally coupled scalar field that satisfies Einstein's
equations $R_{\mu\nu}-\frac{1}{2}R g_{\mu\nu} = T_{\mu\nu}$, where
\begin{equation}
T_{\mu\nu} = \frac{1}{2} \left[  \phi_{,\mu}\phi_{,\nu} - \frac{1}{2}
g_{\mu\nu} \left(  \phi^{,\mu}\phi_{,\mu}+2 \hat{V}(\phi)\right)  \right]  ,
\end{equation}
is
\begin{equation}
\label{phisol}\phi(r) = \frac{2 (k-2)}{\sqrt{3} (k-1)} \ln r
\end{equation}
with a potential given by
\begin{equation}
\hat{V} (\phi) = \pm\frac{6 (4 k-5) e^{-\frac{(k-2) \phi(r)}{\sqrt{3} (k-1)}}%
}{2 k-1} .
\end{equation}
Again the plus sign corresponds to solution \eqref{einfr1}, while the minus is
reserved for \eqref{einfr2}.

Unlike to the previous case, where we have a curvature singularity hidden by
an horizon (thus a black brane/string) for a specific range of the parameter
$k$, here we can see that a singularity is always present at $r=0$. The Ricci
scalar of the resulting spacetimes is
\begin{equation}
R = \pm\frac{6}{(1 - 2 k)^{2} m} r^{-\frac{10 k^{2}-22 k+13}{3 (k-1)^{2}}}
\left[  3 (5 k^{2}-8 k+2) m r^{\frac{(2 k-1) (4 k-5)}{3 (k-1)^{2}}}%
+(k-2)^{2}\right]
\end{equation}
and it can be seen that
\begin{equation}
-\frac{10 k^{2}-22 k+13}{3 (k-1)^{2}}<0, \quad\text{and} \quad\frac{(2 k-1) (4
k-5)}{3 (k-1)^{2}} \leq\frac{10 k^{2}-22 k+13}{3 (k-1)^{2}}%
\end{equation}
hold for all $k \neq1$, with the equality in the latter expression being true
for $k=2$. Thus, we can derive that $R$ is singular at $r=0$ for all values,
but $k=1$ or $k=2$. The first is already excluded from our analysis, while the
latter leads to a constant $R$. However, this does not prevent a curvature
singularity for $k=2$; As we can observe from the Kretschmann scalar
\begin{equation}%
\begin{split}
K =  &  \frac{36}{(1-2 k)^{4} m^{2}} r^{-\frac{2 \left(  10 k^{2}-22
k+13\right)  }{3 (k-1)^{2}}} \Big[ 3 \left(  13 k^{4}-44 k^{3}+54 k^{2}-32
k+10\right)  m^{2} r^{\frac{2 (2 k-1) (4 k-5)}{3 (k-1)^{2}}}\\
&  2 (k-2)^{2} (k (k+2)-5) m r^{\frac{(2 k-1) (4 k-5)}{3 (k-1)^{2}}}+ 23
k^{4}-88 k^{3}+132 k^{2}-100 k+35\Big],
\end{split}
\end{equation}
there exists a curvature singularity for all admissible $k$ in solutions
\eqref{einfr1} and \eqref{einfr2}, even when $k=2$. This latter case
corresponds to having a pure cosmological constant since $\phi=0$ from
\eqref{phisol} and as a result the potential $\hat{V}$ becomes a constant. We can write in this case the line element as
\begin{equation}
\label{knownst}ds^{2} = - \left(  r^{2} - \frac{1}{r}\right)  dt^{2} +
\frac{12}{R} \left(  \frac{1}{r}-r^{2}\right)  ^{-1} dr^{2} + r^{2} \left(
dy^{2}+dz^{2}\right)
\end{equation}
with $\hat{V} = \frac{R}{2}$, where $R$ is the now constant Ricci scalar. Solution
\eqref{knownst} is a known black string solution in the case of a pure
cosmological constant and a special case of a more general solution presented
in \cite{Lemos}. The horizon that hides the singularity is at $r=r_{H}$,
where
\begin{equation}
r_{H} = m^{-\frac{3 (k-1)^{2}}{(2 k-1) (4 k-5)}}
\end{equation}
(of course, for the aforementioned solutions, we consider $m>0$).

As a result of the previous considerations, we can immediately see how the
conformal transformation that takes us from the Jordan to the Einstein frame,
changes the gravitational properties of the solution. A black brane/string is
now admissible for all values of the parameter $k$ for which the solution
holds. Something which did not happen in the $f(R)$ or the O'Hanlon theory.
Although the two frames are conformally equivalent this does not mean that
they are physically equivalent. Solutions can be mapped at a purely
mathematical level, but their study produces in principle different results.

We can also write the basic components of the effective fluid generated by the
scalar field in this case. The energy density $\rho=-T^{0}_{\;0}$ is given by
\begin{equation}
\rho= \pm\frac{3 r^{\frac{-10 k^{2}+22 k-13}{3 (k-1)^{2}}}}{(1-2 k)^{2} m}
\left[  r^{\frac{(2 k-1) (4 k-5)}{3 (k-1)^{2}}} (7 k^{2}-10 k+1) m+(k-2)^{2}
\right]  .
\end{equation}
The radial pressure $p_{r}=T^{1}_{\;1}$ is expressed as
\begin{equation}
p_{r} = \pm\frac{3 r^{\frac{-10 k^{2}+22 k-13}{3 (k-1)^{2}}}}{(1-2 k)^{2} m}
\left[  (k-2)^{2}-9 r^{\frac{(2 k-1) (4 k-5)}{3 (k-1)^{2}}} (k-1)^{2} m
\right]  ,
\end{equation}
while the tangential pressure in this case is $p_{T} = p_{x}=p_{y}=T^{2}%
_{\;2}=T^{3}_{\;3}=-\rho$. Thus, we can see that in this case a ``partial"
equation of state $w_{i} = \frac{p_{i}}{\rho}=-1$ is satisfied in the $x$ and
$y$ directions. It can be easily verified that $\rho$ is positive in the
region $1<k<\frac{5}{4}$, if the radius $r$ satisfies $r>B_{k} r_{H}$, where
\begin{equation}
B_{k} = \left(  -\frac{(k-2)^{2}}{7 k^{2}-10 k+1}\right)  ^{\frac{3 (k-1)^{2}%
}{(2 k-1) (4 k-5)}}.
\end{equation}
In the second region of the solution $k<1$, $k>\frac{5}{4}$, $k\neq{1}{2}$,
the energy density can only be positive in the subsets $\frac{1}{7} \left(
5-3 \sqrt{2}\right)  <k<1$ and $\frac{5}{4}<k<\frac{1}{7} \left(  3 \sqrt
{2}+5\right)  $ given that the radius is again greater than $B_{k} r_{H}$. It
is also now interesting to study the basic thermodynamic quantities to compare
with those derived for \eqref{met.16} and \eqref{met.17} in \cite{nplb}.

\section{Temperature and Entropy}

\label{temp1}

In this section we calculate some basic thermodynamic quantities like the
temperature and the entropy. At first, we can perform a transformation
$t\mapsto\tau$ to bring the metric into an ingoing Eddington-Finkelstein
coordinate system. Thus, if we impose
\begin{equation}
t = \tau\pm\int\!\!\sqrt{\frac{g_{rr}}{-g_{tt}}} dr,
\end{equation}
the line elements \eqref{einfr1} and \eqref{einfr2} are transformed into
\begin{equation}
ds^{2} = \pm r^{2} \left(  1-\frac{r^{-\frac{(2 k-1) (4 k-5)}{3 (k-1)^{2}}}%
}{m}\right)  d\tau^{2} \pm\frac{2(2 k-1)}{3 (k-1)} r^{\frac{(k-2)^{2}}{3
(k-1)^{2}}} d\tau dr + r^{2} \left(  dy^{2} + dz^{2}\right)  .
\end{equation}
with the plus sign corresponding to solution \eqref{einfr1} and the minus to
\eqref{einfr2}. In these coordinates, the surface gravity can either be given
by $\kappa= \Gamma^{\tau}_{\tau\tau}|_{r=r_{H}}$ or - with the help of the
timelike Killing vector $X^{\mu}=\partial_{\tau}$ that exhibits a Killing
horizon at $r=r_{H}$ - by $\kappa= \nabla_{\mu}(X^{\nu}X_{\nu})|_{r=r_{H}} =
-2 \kappa X_{\mu}|_{r=r_{H}}$. In any case, the result is
\begin{equation}
\kappa= \pm\frac{(5-4 k)}{2 (k-1)} m^{\frac{-2 k^{2}+2 k+1}{(2 k-1) (4 k-5)}}
.
\end{equation}
It is rather impressive that the surface gravity is exactly the same with the
one derived in \cite{nplb} for a different set of spacetimes, namely
\eqref{met.15} and \eqref{met.16} that correspond to the relative solutions in
$f(R)=R^{k}$ gravity and which are connected to \eqref{einfr1} and
\eqref{einfr2} through a conformal transformation. We note that once again the
plus sign corresponds to the case $1<k<\frac{5}{4}$, while the minus is for
$k<1$, $k>\frac{5}{4}$, $k\neq\frac{1}{2}$. As a result of $\kappa$ being the same in the two frames, the resulting temperatures at the horizon are also identical. We can see that the latter is given in terms of the single
essential constant of the geometry $m$ and the parameter of the theory $k$ as:
\begin{equation}
\label{temp}T = \frac{\kappa}{2\pi} =
\begin{cases}
\frac{(5-4 k)}{4 \pi(k-1)} m^{\frac{-2 k^{2}+2 k+1}{(2 k-1) (4 k-5)}}, &
1<k<\frac{5}{4}\\
\frac{(4 k-5)}{4 \pi(k-1)} m^{\frac{-2 k^{2}+2 k+1}{(2 k-1) (4 k-5)}}, & k
\in(-\infty, 1) \bigcup(\frac{5}{4},+\infty) - \{\frac{1}{2}\}
\end{cases}
\end{equation}
with each branch corresponding to the appropriate solution for the assumed values of $k$.
The exponent of $m$ in the expressions appearing in \eqref{temp} is negative
in the region $\frac{1}{2}<k<\frac{5}{4}$ and when $k>\frac{1}{2}(1+\sqrt{3})$
or $k<\frac{1}{2}(1-\sqrt{3})$, which are the roots of the polynomial in the
nominator. On the other hand, we have the inverse situation when $\frac{1}%
{2}(1-\sqrt{3})<k<\frac{1}{2}$ and $\frac{5}{4}<k< \frac{1}{2}(1+\sqrt{3})$.

At this point we can write down the area per unit length $\sigma= 2 \pi
r_{H}^{2} = 2 \pi m^{-\frac{6 (k-1)^{2}}{(2 k-1) (4 k-5)}}$. In this manner,
we can define an entropy per unit length as
\begin{equation}
S \sim\frac{\sigma}{ 4} = \frac{\pi}{2} m^{-\frac{6 (k-1)^{2}}{(2 k-1) (4
k-5)}} .
\end{equation}
We can see that the power of $m$ is positive when $k\in(\frac{1}{2},\frac
{5}{4})-\{1\}$, while it is negative for all the other values (excluded of
course $k= \frac{1}{2}, \frac{5}{4}$). The entropy is always well defined and
positive for all admissible values of $k$, in contrast to what happens to the
corresponding solution in the Jordan frame. i.e. the one expressed previously
in the context of the $f(R)$ theory. In the latter, the entropy per unit
length can be found to be \cite{nplb}
\begin{equation}
S_{f(R)} \sim k \left(  \pm\frac{k(4k -5)}{(k-1)(2k-1)} \right)
^{k-1}m^{-\frac{6 (k-1)^{2}}{(2 k-1) (4 k-5)}}.
\end{equation}
The $S_{f(R)}$ of $f(R)=R^{k}$ gravity is positive only in the cases where
$0<k<\frac{1}{2}$ or when $k$ is a negative even integer. This is another
difference in the physics of the system owed to the conformal transformation
linking the two theories. In Einstein's gravity with the minimally coupled
scalar field the solution results in a positive entropy for all values of the
parameter characterizing the potential of the scalar field. We can note
however, that the dependence in what regards the constant $m$ appearing in the
line element remains the same in the two theories.

In both frames the corresponding solution in not asymptotically maximally symmetric. This results in complications in what regards a possible calculation of the mass of the system. Even in the Einstein frame, an indirect computation through the first law of thermodynamics is not trivial due to the contributing work due to the matter source which is characterized by a non constant pressure. This is why we restrict ourselves to carry on the comparison only over the well defined thermodynamic quantities of the temperature and the entropy.

\section{Conclusions}

\label{conc}

In this work we have studied analytical solutions of black branes which are related
through conformal transformations. According to our knowledge this is the
first examples where black brane solutions are compared between the Jordan
and the Einstein frame. At the Jordan frame we consider an analytic solution
which was derived recently for $f\left(  R\right)  $-gravity in the metric
formalism; more specifically in the context of a power law $f\left(  R\right)  =R^{k}$ theory over an
axisymmetric spacetime in vacuum. The conformal equivalent theory in the Einstein frame corresponds to a scalar field with an exponential potential. 

In principle it is expected that not only the physical but also the geometrical properties are different when one passes from one frame to the other. This can be attributed to the fact that conformal transformations do not necessarily correspond to a general coordinate transformation. The latter constitute the gauge freedom transformations of gravitational theories and hence map different line elements describing the same geometry. As a result, when a conformal transformation is performed, the resulting space-time may in principle possess extremely different properties. In our case we where able to witness several striking differences but also a few similarities.

The analytic solution in the Jordan frame was found to describe black branes (solutions with a singularity at $r=0$ that is encompassed by a horizon $r=r_h$) only
for specific values of the free parameter $k$. Moreover, it has been shown that the
entropy of these black brane solutions is positive only for specific values of
$k$, and more specifically when $0<k<\frac{1}{2}$ or when $k$ is a negative,
even integer. The parameter $k$ in the $f\left(  R\right)=R^k$ theory being the power in which the Ricci scalar is raised, while in the Einstein theory is the exponent in the exponential potential characterizing the scalar field. 

In contrast to what happened in the Jordan frame, the conformally equivalent
solution in the Einstein frame was found to describe black branes for all the admissible
real values of the parameter $k$, while the entropy was calculated to be well
defined, i.e. positive, for all the values of this parameter. We can also note that the horizon $r_H$ in the Einstein frame  is shifted $r_H\neq r_h$ when we make the comparison by keeping the same constant $m$ appearing in the two geometries.

A similarity that may be noticed in the two solutions is that  the black objects that they describe are in neither case asymptotically maximally symmetric. Additionally, the surface gravity on the horizon ends up to be the same for both cases. A property which results in the same relation for the thermodynamic temperature as well.

When seeing the solution in the Jordan frame as a result of the O'Hanlon theory, we calculated the energy density and anisotropic pressure of the respective effective fluid. It was interesting to note that, in this situation, by approaching the curvature singularity, the solution looks like the effect of a radiation fluid.  We also derived the relative expressions in context of Einstein's gravity and we investigated for which ranges of the parameter $k$ characterising the two models, the weak energy condition is satisfied.

From all of the above results it follows that the basic physical properties of the aforementioned 
axisymmetric solutions do not survive through the conformal transformation connecting them. Indeed, horizons, the entropy and even singularities change between the two systems. The effect of the latter being the alteration of the conditions over the parameter $k$ needed in order to identify the solution as a black brane. In
a forthcoming work we want to compare the two theories by studying dynamical evolution of test particles among the two different solutions.

\begin{acknowledgments}
This work is financial supported by FONDECYT grant 1150246, CONICYT DPI
20140053 project (AG) and FONDECYT grant 3160121 (AP).
\end{acknowledgments}

\end{document}